\documentclass[fleqn,usenatbib]{mnras}

\usepackage{newtxtext,newtxmath}

\usepackage[T1]{fontenc}

\DeclareRobustCommand{\VAN}[3]{#2}
\let\VANthebibliography\thebibliography
\def\thebibliography{\DeclareRobustCommand{\VAN}[3]{##3}\VANthebibliography}

\usepackage{graphicx}	
\usepackage{amsmath}	
\usepackage[dvipsnames]{xcolor}

\title[Precise Secondary Spectrum Curvatures]{
Interstellar Interferometry:
Precise Curvature Measurement from Pulsar Secondary Spectra}

\author[D. Baker et al.]{
Daniel Baker,$^{1,2}$\thanks{E-mail: dbaker@cita.utoronto.ca}
Walter Brisken,$^{3}$
Marten H. van Kerkwijk,$^{4}$
Robert Main,$^{5}$
Ue-Li Pen,$^{1,4}$
\newauthor
Tim Sprenger,$^{5}$
Olaf Wucknitz$^{5}$
\\
$^{1}$Canadian Institute for Theoretical Astrophysics, University of Toronto, 60 St. George Street, Toronto, ON M5S 3H8, Canada\\
$^{2}$Department of Physics, University of Toronto, 60 St. George Street, Toronto, ON M5S 1A7, Canada\\
$^{3}$National Radio Astronomy Observatory, Socorro, NM 87801, USA\\
$^{4}$ Department of Astronomy and Astrophysics, University of Toronto, 50 St. George Street, Toronto, ON M5S 3H4, Canada\\
$^{5}$Max-Planck-Institut f\"{u}r Radioastronomie, Auf dem H\"{u}gel 69, 53121 Bonn, Germany\\
}

\date{Accepted XXX. Received YYY; in original form ZZZ}

\pubyear{2020}

\begin{document}
\label{firstpage}
\pagerange{\pageref{firstpage}--\pageref{lastpage}}
\maketitle

\begin{abstract}
The parabolic structure of the secondary or conjugate spectra of pulsars is
often the result of isolated one-dimensional (or at least highly anisotropic) lenses in the ISM. The curvature of these features contains information about the velocities of the Earth, ISM, and pulsar along the primary axis of the lens. As a result, measuring variations in the curvature over the course of a year, or the orbital period for pulsars in binaries, can constrain properties of the screen and pulsar. In particular the pulsar distance and orbital inclination for binary systems can be found for multiple screens or systems with prior information on $\sin(i)$. By mapping the conjugate spectra into a space where the main arc and inverted arclets are straight lines, we are able to make use of the full information content from the inverted arclet curvatures, amplitudes, and phases using eigenvectors to uniquely and optimally retrieve phase information. This allows for a higher precision measurement than the standard Hough transform for systems where these features are available. Our technique also directly yields the best fit 1D impulse response function for the interstellar lens given in terms of the Doppler shift, time delay, and magnification of images on the sky as seen from a single observatory. This can be extended for use in holographic imaging of the lens by combining multiple telescopes.  We present examples of this new method for both simulated data and actual observations of PSR B0834+06.
\end{abstract}



\section{Introduction} \label{sec:intro}
Pulsars allowed for the first detection of
gravitational radiation, and provide a promising tool for precise gravitational source astrometry \citep{Boyle2012}. Two current major limitations are the insufficiently accurate distances to pulsars, which would allow for coherent GW measurement; and interstellar propagation effects which degrade the precision of pulsar timing.  Coherent combination of the pulsar intrinsic term doubles the detection sensitivity, and improves the astrometric localization by orders of magnitude, likely localizing GW sources to arc minute precision.
 An improved understanding of the electromagnetic propagation, the positions and magnifications of scattered images, opens up two avenues: interstellar diffraction limited astrometry and descattering of propagation effects.  From the observed intensity field, {\it phase retrieval} aims to recover the linear impulse-response function that, convolved with an emitted pulse yields the observed electric field. This technique is also  called {\it holography} \citep{Walker2005}.  The dynamic wavefield, or its Fourier conjugate the conjugate wavefield, encodes information about the magnification and time delay of images currently contributing to the interference pattern from the plasma lens, and enables its utilization as a giant interstellar interferometer. For a more complete discussion of these objects see Sec.~\ref{sec:lens_model}. \cite{Pen_2014} achieved 50 picoarcsecond relative astrometry using the wavefield decomposition.  To date, the phase retrieval  approaches have been  heuristic, working for special cases using lucky initial guesses and solving non-linear equations with unknown  convergence properties,  potentially leading to local minima  instead of global solutions.  This paper presents a systematic approach: by mapping onto an eigenvalue problem, the existence, uniqueness, optimality and degeneracy of the solutions become explicit.

When pulsar emission is scattered by structures in the interstellar medium (ISM), the resulting scintillation pattern can be used to provide precise measurements of the ISM and the pulsar. For our purposes, we focus on two measurements of interest: pulsar velocities in binary systems, and the structure of the lensing region.\par Since the timescale of scintillation depends on the velocity of our line of sight through the ISM, variations in this timescale can be used to measure velocity changes due to the orbital motion of pulsars in binaries \cite{Lyne1982}. For systems with large orbital velocities, relative to their centre of mass motion, these changes can be of order one and can be measured using the correlation timescale of the scintillation pattern at different orbital phases as per \cite{Rickett_2014}. In order to study systems with less dramatic variation, more precise techniques are required. The discovery of scintillation arcs by \cite{Stinebring2001} presents such an opportunity. In some cases, these features are believed \cite{Walker_2004} (or have been shown \cite{Brisken_2009}) to be the result of linear or highly anisotropic groups of images on the sky at a fixed distance, giving a quadratic relation between the time delay and Doppler shift, and and our analysis will assume this to be the case. The interference of these images produces a main arc, from interference between lensed images and the line of sight image, as well as inverted arclets from the interference of pairs of lensed images. 

The curvature of the main arc, as well as the arclets, is inversely proportional to the square of the effective velocity of the system. For PSR J0437$-$4715, \cite{ReardonPhD} used a Hough transform to measure the power along arcs in the secondary spectrum to measure these curvatures over several years. The annual modulation of the curvature completely encodes the orientation of the screen as well as the effective distance (defined in Sec.~\ref{sec:lens_model}). For binary systems, the variation over the binary period will additionally depend on the pulsar distance, binary inclination, and orbital orientation. From measurements for a single screen this information is encoded in a phase and amplitude and so cannot give all three parameters. Fortunately, $\sin(i)$ is sufficient to restrict parameters to one of two orbits mirrored about the screen on the sky and at a fixed distance. In cases where a second screen can also be detected, which \cite{Putney_2006} observe in multiple systems, modeling the evolution of both screens can resolve the degeneracies and yield both the distance and inclination. Broadly, each screen has five measurable parameters, an amplitude and phase of modulation for both the Earth and pulsar orbits and a constant from the proper motion and screen velocity. However, each screen introduces only three new unknown parameters: screen distance, orientation and velocity. In \cite{ReardonPhD}, the nearly parallel screens prevent measurement of the distance. Fortunately, the detection of additional screens has allowed for distance measurement with order ten per cent uncertainty (private communication). Improvements in the curvature measurements for the individual observations should allow for further improvement.  For isolated pulsars similar techniques, using only curvature modulations from the Earth's orbit, can help constrain screen distances and orientations.

In this paper, we present a technique for measuring curvatures with greater precision by using not just the power along the main arc, but the phases and amplitudes of the main arc and inverted arclets. This is achieved by mapping the arcs into a space where they are represented by linear features proposed by \cite{Sprenger_2020},dubbed the $\theta-\theta$ transformation, and discussed here in Section~\ref{sec:th-th}. In Section~\ref{sec:simulated_eta}, we present a sample application of this method to some simulated data to show how it can improve precision on curvature measurements.

An additional advantage of the $\theta-\theta$ transformation is that it can be used to solve the phase retrieval problem for pulsar scintillometry. Though much progress has been made in studying pulsar scintillometry through the dynamic and secondary spectra, the loss of phase information in the formation of the dynamic spectrum from the square modulus of the electric field imposes serious restrictions. The problem of phase retrieving this phase information in the context of pulsar scintillometry was first discussed by \cite{Walker2005}.  By reconstructing the phase of the wavefield we gain information about the locations and signal delays of the individual images causing scintillation. Apart from giving information about the lensing structure in the ISM,  it is hoped that this information can be used to better understand changes in the measured time of arrivals caused by time evolution in the scattering medium between epochs \cite{Walker2005}. In studying seven years of PSR J0613-0200 data, \cite{Main_2020} find the apparent strain due to variations in scattering to be $h \approx 10^{-15}$ at 15 nHz, where \cite{Aggarwal_2018} had previously reported an apparent signal due to an unmodelled signal in this pulsar in the NANOgrav 9-year dataset. Though below the current single pulsar limit of $9.7 \times 10^{-15}$, they argue that as PTA upper limits improve these effects may limit precision.

In Section~\ref{sec:lens_model}, we discuss a simple one-dimensional model of scintillation. Section~\ref{sec:th-th} shows how $\theta-\theta$ transformations can be used for phase retrieval when only the dynamic spectrum is available, and in Section~\ref{sec:PhaseRet} we demonstrate its effectiveness on simulated data.

In Section~\ref{sec:B0834} we apply these methods to an observation of PSR B0834+06 from \cite{Brisken_2009} using Arecibo. This data shows clear inverted arclets, which lend themselves well to this style of analysis. The secondary spectrum also includes clear deviations from the one-dimensional model, as seen by collection of power offset from the main arc near $1 \text{ms}$. We show how modelling the one-dimensional structure responsible for the main arc and arclets can be used to probe this region using phase retrieval.

\section{Theory of the 1D Lens}\label{sec:lens_model}
\begin{figure*}
    \centering
    \includegraphics{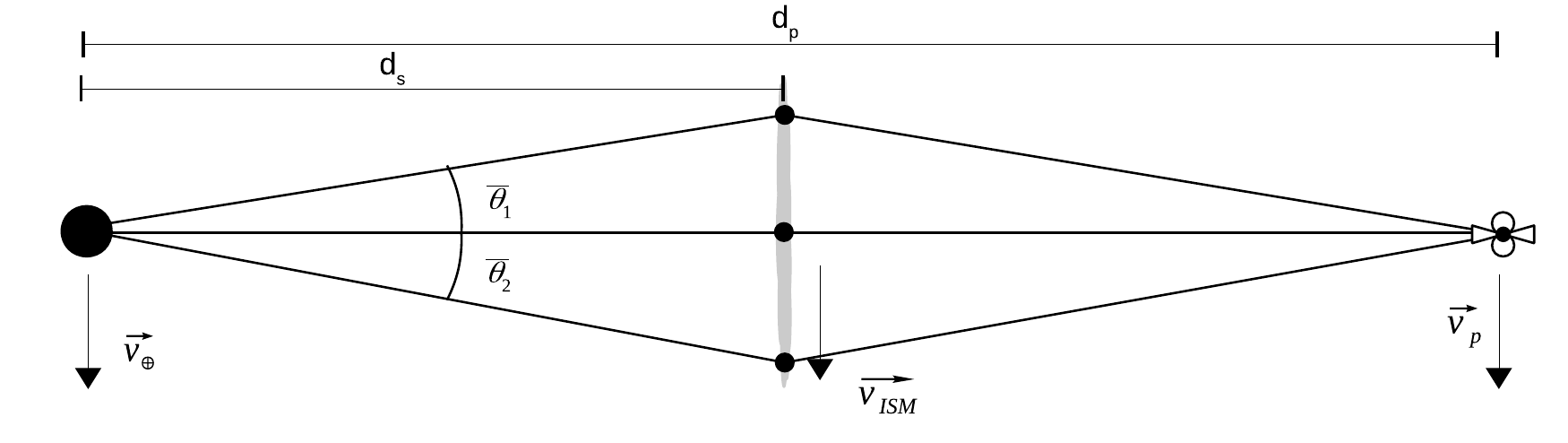}
    \caption{Schematic of one-dimensional lensing with three images (including the direct line of sight). Each image has a time delay and Doppler shift determined by $\Bar{\theta}$, and a magnification $\Bar{\mu}$ determined by local properties of the ISM. Angles have been exaggerated for this diagram, and in reality are expected to be on mas scales.}
    \label{fig:geometry_scheme}
\end{figure*}

Since our method for phase retrieval and curvature measurement is based on the assumption of a one-dimensional screen, we briefly introduce how such a structure leads to the observed spectra and how it relates to the physical properties of interest. For a pulsar located at a distance $d_p$, we assume some lensing structure in the ISM, localised at some distance $d_s$, producing a line of images of the pulsar on the sky. For any given image along the line with angular offset $\Bar{\theta}$, the pulsar signal will experience a geometric delay and Doppler shift, at observational wavelength $\lambda$, of
\begin{eqnarray}
    \tau\left(\Bar{\theta}\right) &=& \frac{\Bar{\theta}^2 d_{\text{eff}}}{2c}\\
    f_D\left(\Bar{\theta}\right) &=&  \frac{v_{\parallel} \Bar{\theta}}{\lambda}
\end{eqnarray}
where
\begin{eqnarray}
    d_{\text{eff}} &=& \frac{d_p d_s}{d_p-d_s}\\
    v_{\parallel} &=& \left(\Vec{v}_{\oplus} + \Vec{v}_p \frac{d_s}{d_p-d_s} - \Vec{v}_{\text{ISM}}\frac{d_p}{d_p-d_s}\right)\cdot \vec{n}
\end{eqnarray}
for pulsar, Earth and ISM velocities $\Vec{v}_{p}$,$\Vec{v}_{\oplus}$ and $\Vec{v}_{\text{ISM}}$ respectively; and $\vec{n}$ is the unit vector from the line of sight to the lensed image on the sky. It is clear that for a collection of images their time delay depends quadratically on the Doppler shift as
\begin{equation}
    \tau = \eta f_D^2
    \label{eq:eta_def}
\end{equation}
where $\eta$ is the curvature of the observed arc in the secondary spectrum and is given by
\begin{equation}
    \eta = \frac{d_{\text{eff}} \lambda^2}{2cv_\parallel^2}
\end{equation}
Since the typical time delays and Doppler shifts in L, 1-2 GHz, and P, 230-470 MHz, bands are on the order $\rm\mu \text{s}$ and $\rm \text{mHz}$, curvatures are naturally given in $\rm\mu{}s~ \text{mHz}^{-2}$. However, in keeping with previous works we will be quoting curvature in $\rm \text{s}^3$, which fortunately converts as $1~ \rm\mu{}\text{s}~\text{mHz}^{-2} = 1~\text{s}^3$.
The geometry of such a system with three images, including the unlensed image, is shown in Fig.\ref{fig:geometry_scheme}.

In addition to these geometric effects, each image also has a complex magnification $\Bar{\mu}$ determined by the local properties of the lensing medium. We will call the magnification as a function of $\theta$ the impulse response, as it shows the response of the lensing medium to a single coherent source.

The interference of these images produces the observed electric field at Earth. The average intensity of this field, usually over several pulses in time, as a function of time and radio frequency is known as the dynamic spectrum $I(t,\nu)$ whose Fourier transform $\Tilde{I}(f_D,\tau)$ is called the conjugate spectrum. Underlying the dynamic spectrum is the complex wavefield $U(t,\nu)$ whose square modulus is the dynamic spectrum. For consistency, we call the 2-D Fourier transform of this object the conjugate wavefield $\Tilde{U}(f_D,\tau)$. For a one dimensional lens, the conjugate wavefield will be zero everywhere except along the parabola given by Eq. ~\ref{eq:eta_def} where it will be equal to the complex magnification of each image. The wavefield can also be related to the scattering time function by a Fourier transform along frequency, giving the scattering time function for each time bin.  Since the intensity is the amplitude squared of the wavefield, the conjugate spectrum must be the convolution of the conjugate wavefield with its complex conjugate transpose. For a one dimensional lens this will result in the familiar arc and inverted arclets. For more complicated lensing structures the resulting secondary spectrum may be quite different. However, if there exists a dominant linear feature on the sky, the spectrum may still be approximated as the convolution of its parabola with the entire conjugate wavefield. An example of this behaviour is seen in the millisecond feature of Fig.\ref{fig:GB057_wavefield} and discussed in Section~\ref{sec:B0834}. 

\section{$\theta-\theta$ Transformation}\label{sec:th-th}
Since, or our analysis, the parabolic features in the secondary spectrum arise from one-dimensional structures on the sky, there exists a space in which the features are linear; and so easier to work with. Mapping into this space was first proposed by \cite{Sprenger_2020} via their $\theta-\theta$ transformation, which we present here in units of Doppler shift as opposed to sky position
\begin{eqnarray}
    f_D &=& \theta_1 - \theta_2\\
    \tau &=& \eta \left(\theta_1^2 - \theta_2^2\right) = \eta f_D(\theta_1+\theta_2)\\
    \theta_1 &=& \left(\frac{\tau}{\eta f_D} +f_D\right)/2\\
    \theta_2 &=& \left(\frac{\tau}{\eta f_D} -f_D\right)/2
\end{eqnarray}
Where $f_D$ and $\tau$ are the Doppler shift and delay of points in the conjugate spectrum respectively, $\eta$ is the curvature of the main arc, and $\theta_1$ and $\theta_2$ are scaled sky coordinates at which the images would interfere at the given $\tau$ and $f_D$. Our scaled $\theta$ can be converted into angles on the sky with
\begin{equation}
    \Bar{\theta} = - \frac{\lambda  \theta}{v_{\parallel}}
\end{equation}
We define the flux preserving map on the conjugate spectrum as
\begin{equation}
    \Tilde{I}\left(\theta_1,\theta_2;\eta\right) = \Tilde{I}\left( f_D\left(\theta_1,\theta_2\right),\tau\left(\theta_1,\theta_2; \eta\right)\right) \sqrt{|2\eta\left(\theta_1-\theta_2\right)|}
\end{equation}
By mapping the conjugate spectrum instead of the secondary spectrum, we preserve the phase information which allows for a coherent curvature search as well as the possibility of recovering the wavefield.\par
To see how this is achieved, we consider a collection of images along a line on the sky with angular offsets from the line of sight to the pulsar given by $\Bar{\theta_i}$, where the bar denotes the true angular position of the images as opposed to their curvature-dependent position in a $\theta-\theta$ map. At a fixed frequency, each of these images have both a magnification and phase rotation that are combined into the complex magnification $\Bar{\mu}_i$. Together, these magnifications give the magnification vector $\vec{\mu}$. A simple schematic of the geometry is shown in Fig.\ref{fig:geometry_scheme}.
If we define a grid in the corresponding scaled $\theta$; then transforming into $\theta-\theta$ space, with the correct curvature, gives

\begin{equation}
    \Tilde{I}_{i,j} = \Tilde{I}(\theta_i,\theta_j) = \Bar{\mu}_i \Bar{\mu}_j^*
\end{equation}
and so we can express the $\theta-\theta$ spectrum as the outer product of the magnification vector with itself. This vector is then the only eigenvector of this matrix with a nonzero eigenvalue.
In order to find the correct curvature and magnification vector, we fit our data by minimizing the $\chi^2$ defined by
\begin{equation}
    \chi^2 = \sum_{i,j}\frac{\lvert\Tilde{I}_{i,j}^{\eta}-\mu_i\mu_j^*\rvert^2}{\sigma_{i,j}^2}
    \label{eq:chisq}
\end{equation}
where $\sigma^2_{i,j}$ gives the noise as each point in the $\theta-\theta$ spectrum and $\Tilde{I}_{i,j}^{\eta}$ is the $\theta-\theta$ spectrum for curvature $\eta$. For a fixed $\eta$, the local minima satisfy
\begin{equation}
    \frac{\partial \chi^2}{\partial \mu_i} = -\sum_{j}\frac{\mu_j^*}{\sigma_{i,j}^2}\left(\Tilde{I}_{i,j}^{\eta*} +\Tilde{I}_{j,i}^{\eta}-2\mu_i^*\mu_j\right)=0
    \label{eq:minima}
\end{equation}
Since the dynamic spectrum is real, the conjugate spectrum and by extension the $\theta-\theta$ spectrum are Hermitian. Under the assumption that the noise level is constant for all points, Equation~\ref{eq:minima} simplifies to

\begin{equation}
    \sum_{j}\mu_j\Tilde{I}_{i,j}^{\eta} = \mu_i\sum_{j}\lvert\mu_j\rvert^2
\end{equation}
or

\begin{equation}
    \Tilde{I}^{\eta}\mu=\lvert\mu\rvert^2\mu.
\end{equation}
Hence, minima correspond to eigenvectors that are scaled such that their norm squared equals the corresponding eigenvalues. To determine the global minimum, we rewrite Equation~\ref{eq:chisq} as
\begin{equation}
     \chi^2 = \frac{1}{\sigma^2}\sum_{i,j}\lvert \Tilde{I}_{i,j}^{\eta}\rvert^2 +\lvert\mu_i\rvert^2\lvert\mu_j\rvert^2-\mu_i^*\Tilde{I}_{i,j}^{\eta}\mu_j-\mu_i \Tilde{I}_{i,j}^{\eta,*}\mu_j^*
     \label{eq:chisq_global}
\end{equation}
and so, using that the local minima all satisfy the eigenvector condition above,
\begin{equation}
     \chi^2 = \frac{1}{\sigma^2}\left(\sum_{i,j}\lvert \Tilde{I}_{i,j}^{\eta}\rvert^2 - \lambda_{\eta,n}
     ^2\right)
\end{equation}
where $\lambda_{\eta,n}$ is the $\text{n}^{\text{th}}$ eigenvalue for curvature $\eta$.

Therefore, the best fit magnification vector at a given curvature corresponds to the
largest eigenvalue, and the best fit curvature is the one whose $\theta-\theta$ spectrum has the largest eigenvalue. Since our model takes the outer product of the magnification vector, the solution is not unique under phase rotations which will not impact our curvature fit but is addressed in Section~\ref{sec:PhaseRet} when determining the wavefield.
It should be noted that we have assumed that the noise in $\theta-\theta$ space will be white and of constant variance . In general, this will not be the case as the normalization of the $\theta-\theta$ matrix will scale the stationary noise of the conjugate spectrum and the correlation of points will depend on how the spectrum is sampled. 

\section{Curvature Fitting on Simulated Data}\label{sec:simulated_eta}
To present a proof of concept of the method, as well as the details of the procedure, we simulate the scintillation pattern of a one-dimensional screen. The simulation generates a Gaussian distribution of image positions along a line. Each of them is treated as a stationary phase point, where the combination of dispersive and geometric delays remains constant over some region on the screen at a reference frequency, with a random magnification and phase. For each time in the simulated dynamic spectrum, the phase evolution for each image over frequency is determined and then combined. As time progresses and the pulsar moves, the relative geometric delays of the images change producing the changing field. The dynamic spectrum is then calculated from the average amplitude squared of the electric field and Gaussian noise is added to each point.

\begin{figure}
    \centering
    \includegraphics{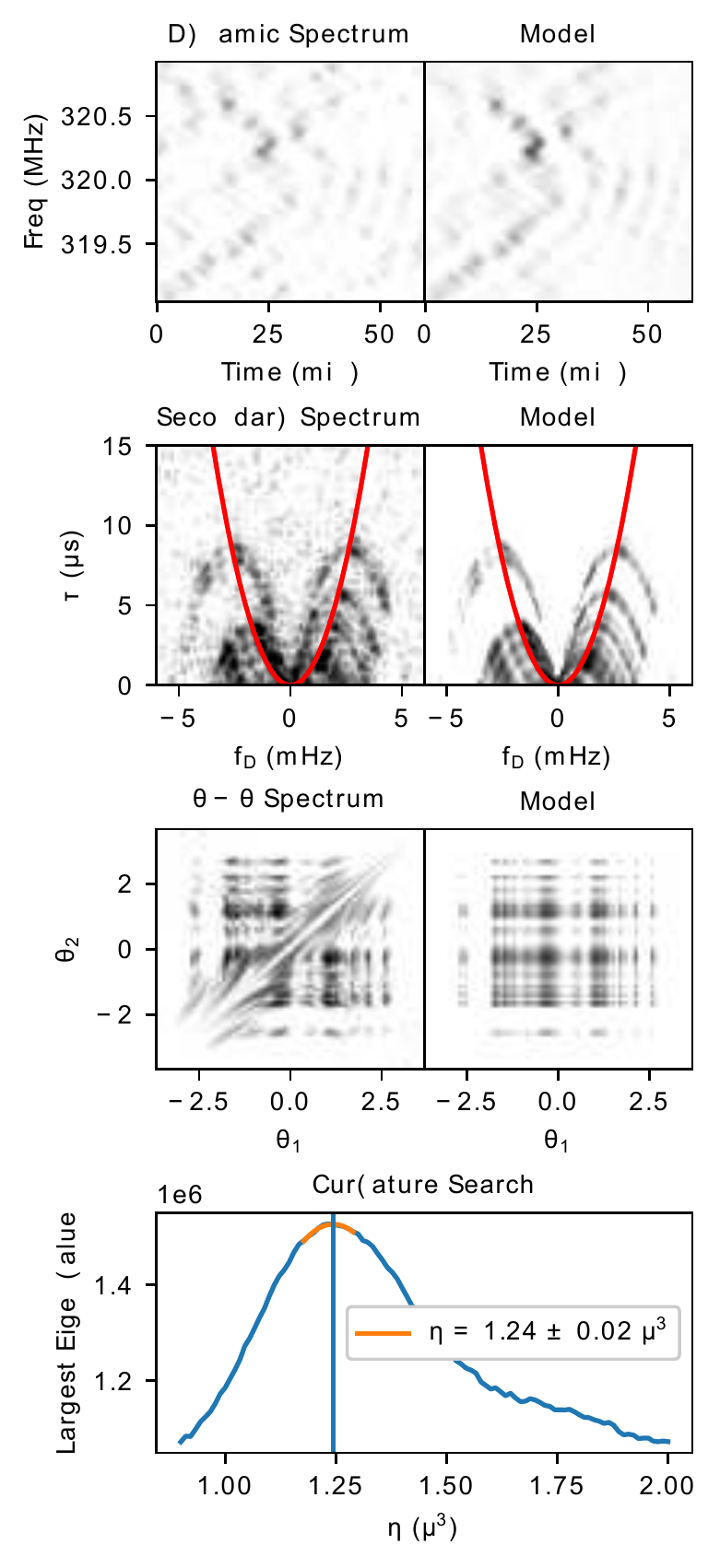}
    \caption{Curvature measurement on a single chunk of the simulated data. The shown $\theta-\theta$ spectra are for the best fit curvature, and the models are generated from its dominant eigenvector. All models use the same colour scale as their respective data. $\theta-\theta$ spectra have been flipped vertically for readability, but the $\theta_1=\theta_2$ diagonal is truly the main diagonal of the matrix. The red line on the secondary spectra plots shows the best fit curvature.}
    \label{fig:simulated_single_curvature}
\end{figure}

Our simulation uses the parameters given in Table~\ref{tab:sim_params}, where the expected curvature at $320~\text{MHz}$ is $1.244~\text{s}^3$. A one hour $2.5~\text{MHz}$ chunk of the simulation, ${1/16}^{th}$ of the band and ${1/4}^{th}$ the observation time, is shown in Fig.~\ref{fig:simulated_single_curvature}.
\begin{table}
\caption{Parameter choices for the simulated dynamic spectrum.}
\centering
\begin{tabular}{||c|c||}
    \hline
    \multicolumn{2}{|c|}{Simulation Parameters} \\
    \hline
    Pulsar Distance & $1.58~\text{kpc}$ \\
    Screen Distance & $0.79~\text{kpc}$ \\
    Projected Proper Motion & $31.99~\text{mas}~\text{yr}^{-1}$ \\
    Observing Band & $310.5~\text{MHz}$ - $340.5~\text{MHz}$ \\
    Number of Channels & 1024 \\
    Observation Length & $4~\text{h}$ \\
    Time Bins & 600 \\
    \hline
\end{tabular}
\label{tab:sim_params}
\end{table}

Since the curvature of the arc evolves in frequency and the relative phases of the images evolve over time, a coherent curvature fit requires that we use only a small region of the dynamic spectrum such that the conjugate wavefield, or equivalently the scattering time function of the lens, remains relatively constant. Using smaller chunk sizes also reduces the resolution of the secondary spectrum. For small deviations from a one dimensional line of images, choosing a coarse enough resolution will cause the images to appear one dimensional and subject to our analysis. For such a chunk, curvature is measured as follows:
 \begin{enumerate}
     \item The mean subtracted dynamic spectrum is zero padded to account for the assumption of periodicity in the FFT and increase the resolution of the conjugate spectrum.
     \item The conjugate spectrum is generated with an FFT.
     \item Determine the grid on which $\theta-\theta$ spectra will be generated. The extent of the grid is determined by the position of the peak of the most outlying arclet of interest, while the resolution is chosen in order to oversample the secondary spectrum.
     \item Generate the $\theta-\theta$ spectrum on the fixed grid of $\theta-\theta$ space for a given curvature.
     \item Perform an eigenvector decomposition and save the largest eigenvalue.
     \item Repeat steps (iv) and (v) over a range of curvatures.
     \item Fit a parabola to the peak of eigenvalue vs curvature.In this work, we fit this parabola using all curvatures within approximately ten percent of the peak. We find that asymmetries in the eigenvalue vs curvature curve may bias the fit if too large a region is used.
 \end{enumerate}
 The results of this approach on a single chunk can be seen in Fig.~\ref{fig:simulated_single_curvature}. Though fitting only requires finding the dominant eigenvalue, we include models for the $\theta-\theta$, secondary, and dynamic spectra as a sanity check. These models are created by taking the outer product of the dominant eigenvalue for the best fit curvature transformation with itself, scaled by the eigenvalue. Inverting the transformation gives us the conjugate spectrum, from which the dynamic and secondary spectra are generated in the usual fashion. Since the $\theta-\theta$ model is built under the assumption the wavefield remains constant, that is to say the locations and magnifications of images, over the chunk being analyzed, examining the model dynamic spectrum can help determine the appropriate  bandwidth and duration for our chunks. When too large a chunk is selected, the model tends to accurately reproduce the brightest scintles and become less accurate further away. To see why, we consider dividing the chunk into smaller regions and expressing it as the sum of these regions zero padded to the original size. The $\theta-\theta$ matrix of the whole chunk is then the sum of the $\theta-\theta$ matrices of the smaller regions. Any region with bright scintles will have a larger contribution to this combined matrix and force the result towards its response function. If the response function evolves over the chunk, then other sections will be less well recovered.

\begin{figure}
    \centering
    \includegraphics{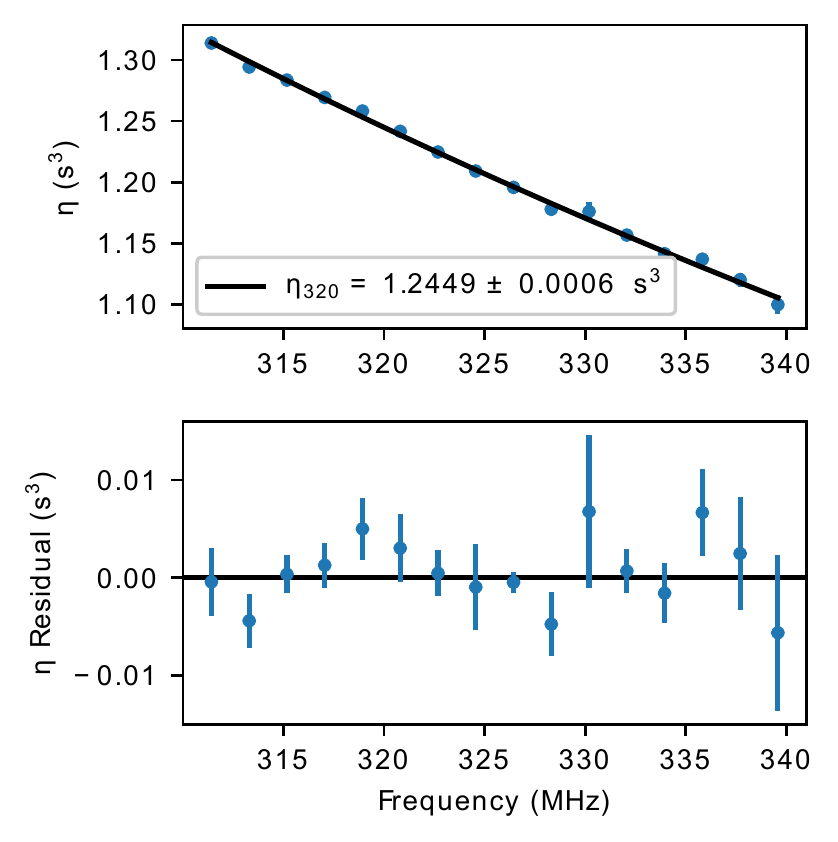}
    \caption{Measured curvatures for the sixteen frequency chunks of the simulated data set. The top panel shows the average curvature at each frequency along with the best fit model. The lower panel shows the residuals.}
    \label{fig:sim_curve_evo}
\end{figure}

Since each $\theta-\theta$ curvature fit is for only a portion of the dynamic spectrum, we can combine multiple chunks to further improve precision.From Equation~\ref{eq:eta_def}, $\eta\propto f^2$ and so we can fit the curvatures to some reference frequency as
\begin{equation}
    \eta = \eta_{\rm{ref}}\left(\frac{f_{\rm{ref}}}{f}\right)^2
\end{equation}{}
In order to estimate the error on the measured curvatures we take the mean and standard error from the curvatures of all chunks at the same frequency. In the case of our simulated data, we have four measurements at each curvature.
Fitting to our reference frequency of $320~\text{MHz}$ (Fig.~\ref{fig:sim_curve_evo}) gives $\eta_{320} = 1.2449 \pm 0.0007~\rm \text{s}^3$ which differs from our expected value of $1.244~\rm \text{s}^3$ by less than a tenth of a percent.

Using a Hough Transform method, as per \cite{ReardonPhD}, applied to the same data yields $\eta_{320} = 2.3 \pm .03~\rm \text{s}^3$. The likely cause of the larger bias here is the asymmetry of the lens. As seen in Fig~\ref{fig:simulated_single_curvature}, there is additional power inside the best fit arc for arclets at negative $f_D$, and outside the arc for those with positive $f_D$. This indicates that there is more power in the images with negative $f_D$ in the conjugate wavefield. Since the inside left of the parabola in the secondary spectrum is due to the interference of these brighter images with each other, it will dominate over features outside the arc on the right. As a result, there is a tendency to pull the arc inward and measure a higher curvature. For cases where the arc is more symmetric or narrower the Hough transform method has been seen to accurately measure curvatures.
Our new coherent curvature fit reduces the curvature estimation error by more than an order of magnitude, while also reducing the bias due to asymmetric arcs, which will aid in future measurements of pulsar physical parameters such as mass and distance using the methods described in \cite{reardon2020precision} and \cite{van_Kerkwijk_2011}.
{}

\section{Phase Retrieval on Simulated Data}
\label{sec:PhaseRet}
Once we have a measurement of the reference curvature, we can recreate the wavefield phase. Phase retrieval is performed on the same sized chunks of data as for curvature fitting. For each piece of the data, the best fit curvature for the central frequency from the curvature model is used to generate the $\theta-\theta$ matrix for which the response vector is determined using eigenvalue decomposition. The wavefield is then determined by using the inverse $\theta-\theta$ map on the response vector with $\theta_2=0$ to place images on the main parabola. However, since the eigenvectors are not unique under a constant phase rotation the recovered phases of these chunks cannot be directly combined. In order to solve this problem, we perform the reconstruction on overlapping chunks and rotate the phases of the recovered wavefields to agree within the overlapping regions, we refer to this as the 'mosaic'. For simplicity, we choose our chunks such that they overlap halfway in time and frequency with the adjacent chunks. Starting from the first chunk in time and frequency, we first apply a Hann window to each chunk, since the edges of the recovery are less reliable. The phase correction applied to the $\rm n^{th}$ chunk is given by
\begin{equation}
    \phi_n = \arg\left(\left<E E_n^*\right>\right)
\end{equation}
where $E$ is the current estimate for the wavefield overlapping with the chunk, and $E_n$ is the windowed wavefield of the chunk. The windowing gives a higher weighting to points away from the edges of chunks to reduce edge effects. We then update $E$ by adding $E_n e^{i \phi_n}$ to the appropriate region. Because of the Hann window, the final field is the weighted mean of rotated chunks.

\begin{figure}
    \centering
    \includegraphics{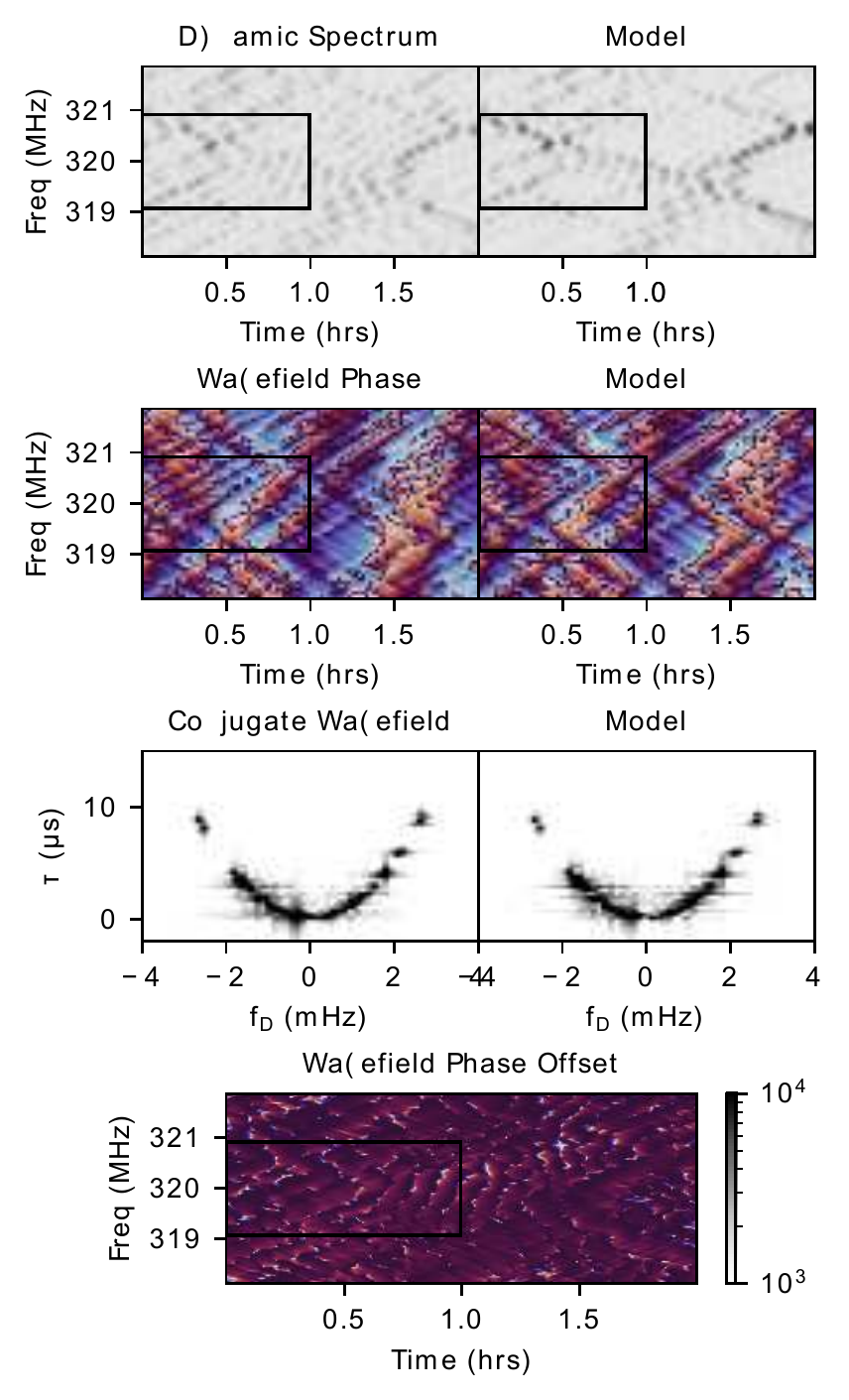}
    \caption{A comparison of the true wavefield in the simulations to the recovered wavefield. The portion of the dynamic spectra and wavefield shown is a mosaic of several chunks as described in the text. The colorbars are the same for the data and model in each row. A constant phase correction has been applied to the model phase for display purposes. The black rectangle marks the location of the chunk used in Fig.\ref{fig:simulated_single_curvature}.}
    \label{fig:sim_recon_chunk}
\end{figure}

The results of this approach are shown in Fig.~\ref{fig:sim_recon_chunk}. The model dynamic spectrum captures the structure of the data nicely, but due to the unknown phase rotation of the first chunk there is a constant shift in the phase model. For the final conjugate wavefield model, we use that the square root of the  dynamic spectrum gives another measurement of the wavefield amplitude, and reapply those amplitudes to our recovered field. The noise properties of the recovered amplitudes and phases, as well as their dependence on measurement noise, are left for a future work.
{}

\section{PSR B0834+06}\label{sec:B0834}
\begin{figure}
    \centering
    \includegraphics{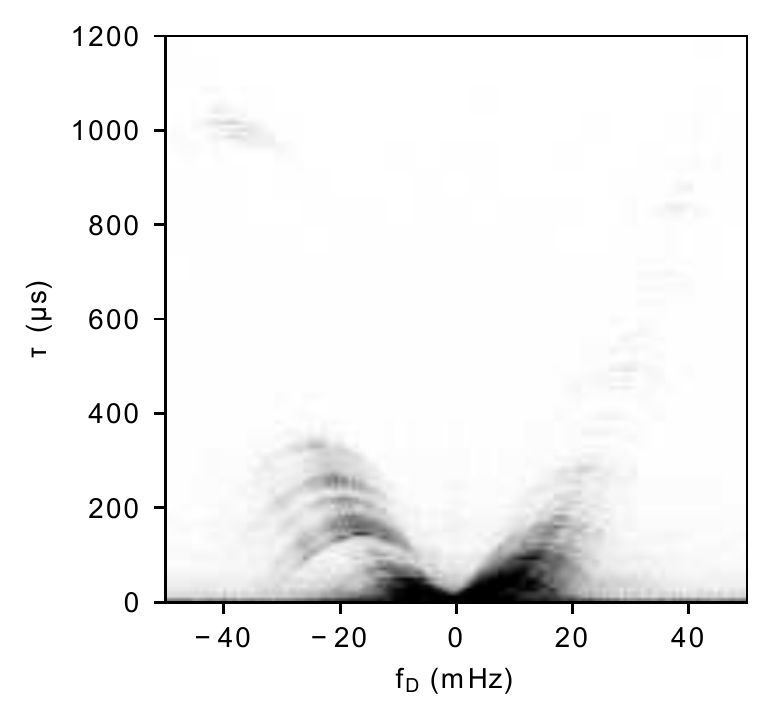}
    \caption{The secondary spectrum of PSR B0834+06 using the bottom $4~\rm \text{MHz}$ of the Arecibo observation. The millisecond feature at negative Doppler frequencies, and clear inverted arclets make this spectrum an interesting test case for $\theta - \theta$ methods}
    \label{fig:GB057_SS}
\end{figure}

\begin{figure}
    \centering
    \includegraphics{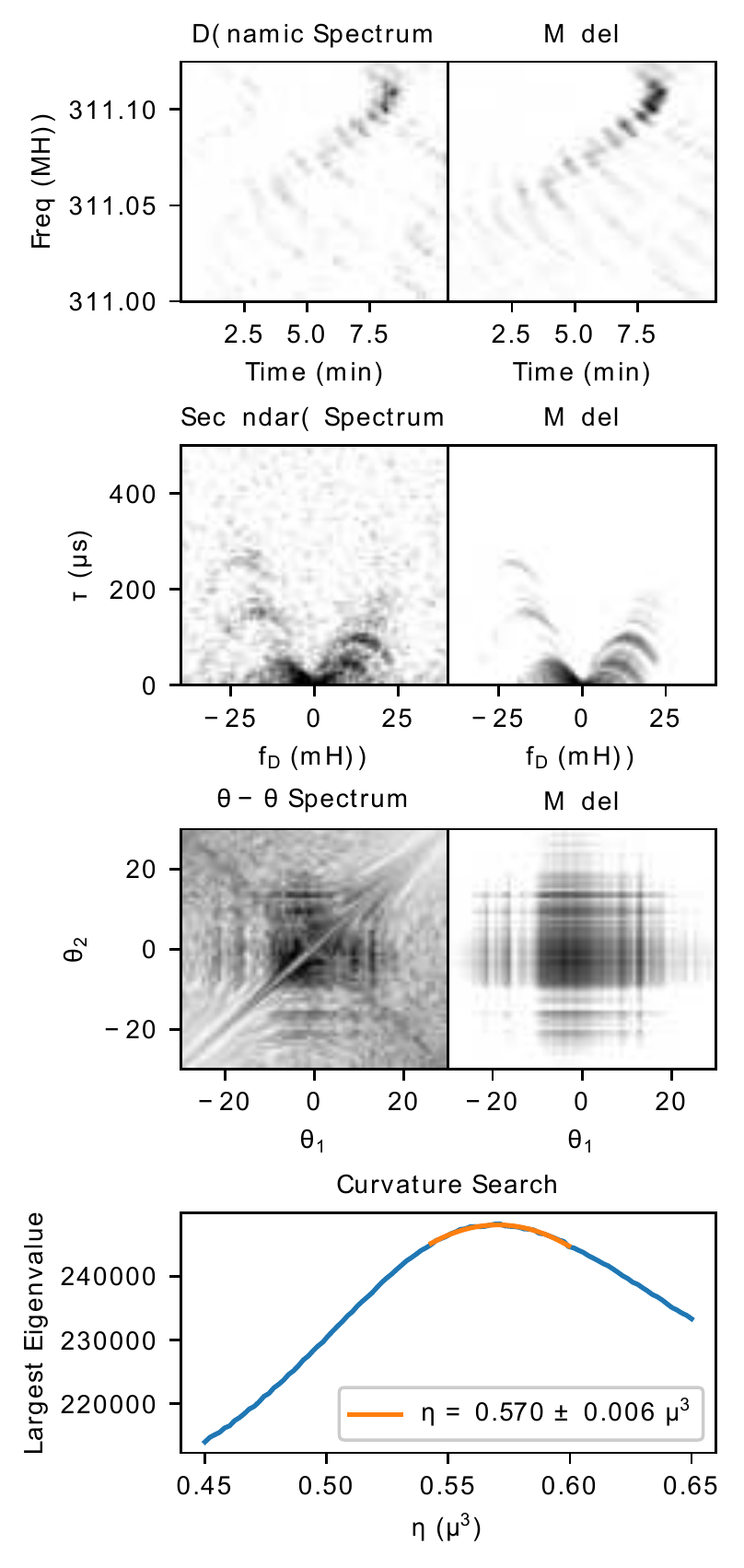}
    \caption{Curvature measurement on a single chunk of the B0834 dynamic spectrum. The $\theta-\theta$ spectrum and model are shown for the best fit curvature. All models are shown using the same colour map as the original data.}
    \label{fig:GB057_single}
\end{figure}

\begin{figure}
    \centering
    \includegraphics{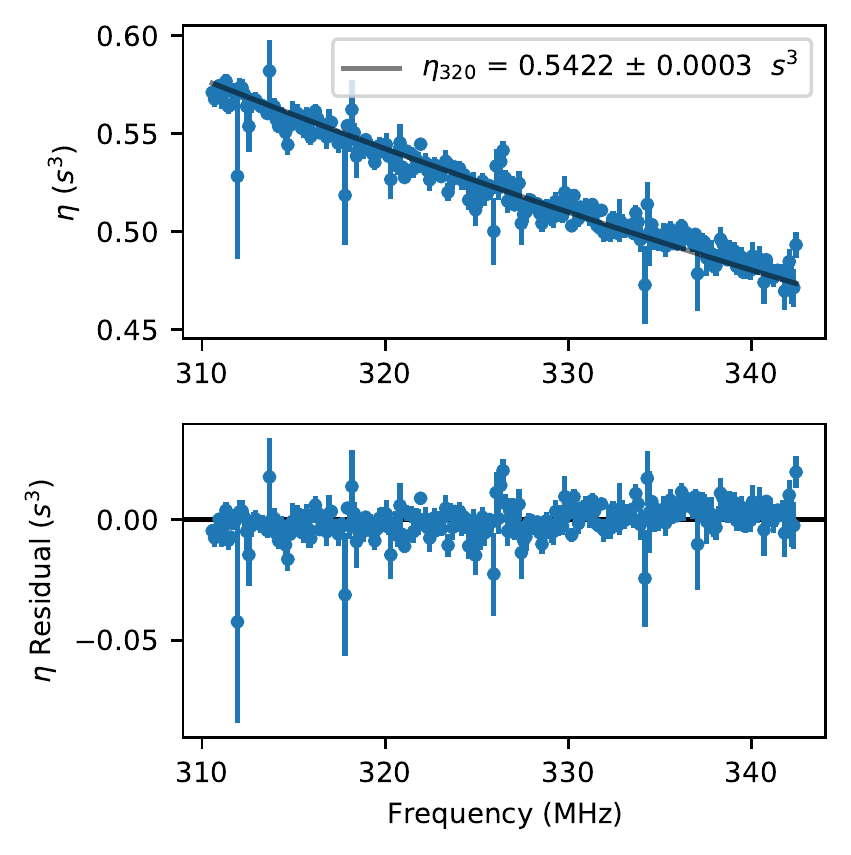}
    \caption{Curvature evolution of B0834+06 data. The top panel shows average curvature of all chunks at the same frequency along with the best fit model. The lower planel shows the residuals of this fit. The reduced chi-squared for the fit is $\chi_{\text{red}} = 0.92$ with 255 degrees of freedom.}
    \label{fig:GB057_curve_evo}
\end{figure}

A useful test of our methods on real data, from which we can also investigate interesting science questions, are the 2005 observations of PSR B0834+06 from \cite{Brisken_2009}  in a $32~\rm \text{MHz}$ band centered at $316.5~\rm \text{MHz}$ for approximately $110~\rm min$. For this test, we restrict ourselves to the Arecibo data, though an extension of $\theta-\theta$ transformation technique for VLBI is under development. The secondary spectrum from the lowest $4~\rm \text{MHz}$ of the observation is shown in Fig.~\ref{fig:GB057_SS}. The presence of clear inverted arclets makes this an excellent candidate for $\theta-\theta$ mapping, while the island of power near $1~\rm \text{ms}$ allows us to examine how non linear features impact recovery. \cite{Brisken_2009} show that this feature does not lie on the main parabola, and can be mapped to a different linear screen on the sky. Since the eigenvector decomposition $\theta-\theta$ matrices assumes a single curvature for the screen, we must discount this feature from our initial model. As most of the power in the main arc is below $512~\rm\mu \text{s}$, we remove the island by rebinning the dynamic spectrum by a factor of four in frequency by averaging each consecutive group of four channels. As before, we divide the dynamic spectrum into sections and perform a curvature measurement on each one, with a characteristic result shown in Fig.~\ref{fig:GB057_single}. Using only $0.125~\rm \text{MHz}$ of the band and $10.5~\rm \text{min}$ of the data, we are able to measure a curvature to almost one per cent. For this chunk size, we can make more than 2500 curvature measurements from this dataset. These measurements are treated as independent as this approach yields a reasonable error estimate for the simulated data. The results of these individual measurements are combined in Fig.~\ref{fig:GB057_curve_evo} by averaging all curvature measurements for chunks at the same frequency and using the standard error, fitting their curvature evolution gives $\eta_{320} = 0.5422 \pm 0.0003~\rm \text{s}^3$ with a reduced chi-squared value of $\chi_{\text{red}} = 0.92$ with 255 degrees of freedom.


\begin{figure*}
    \centering
    \includegraphics{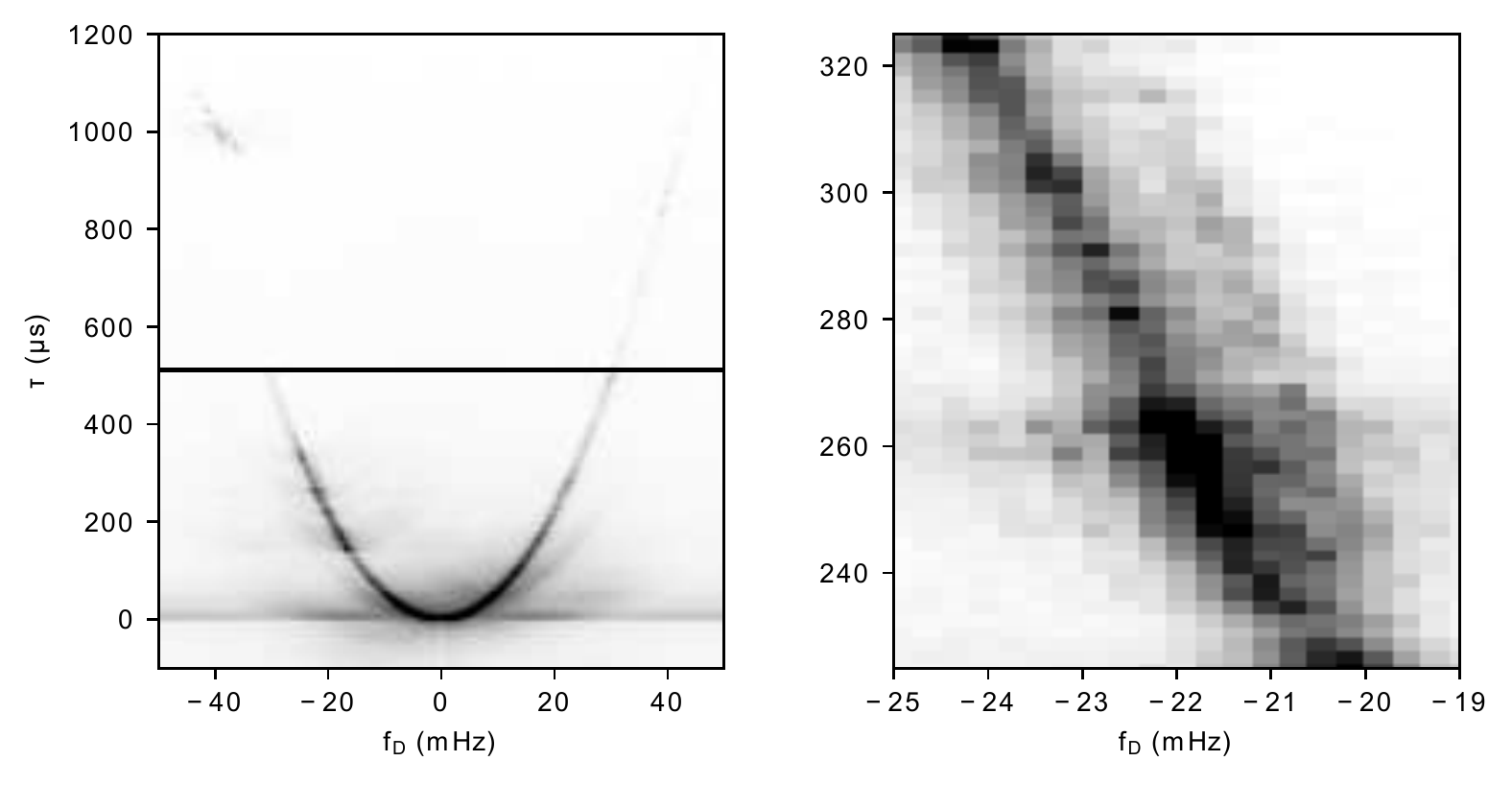}
    \caption{The recovered conjugate wavefield for B0834+06. The figure on the left shows the whole conjugate wavefield, including the millisecond feature, while the figure on the right shows a zoomed in version at $300\mu s$ where a second group of images can be seen coming off the main arc. Since the dynamic spectrum was rebinned before performing $\theta-\theta$ analysis, only the region below $512\mu s$, marked by a black line on the left, was originally modelled. Interpolation to the original resolution, and applying amplitudes from the square root of the dynamic spectrum, allows us to see nonlinear features and those above $512\mu s$.
    }
    \label{fig:GB057_wavefield}
\end{figure*}
\begin{figure*}
    \centering
    \includegraphics{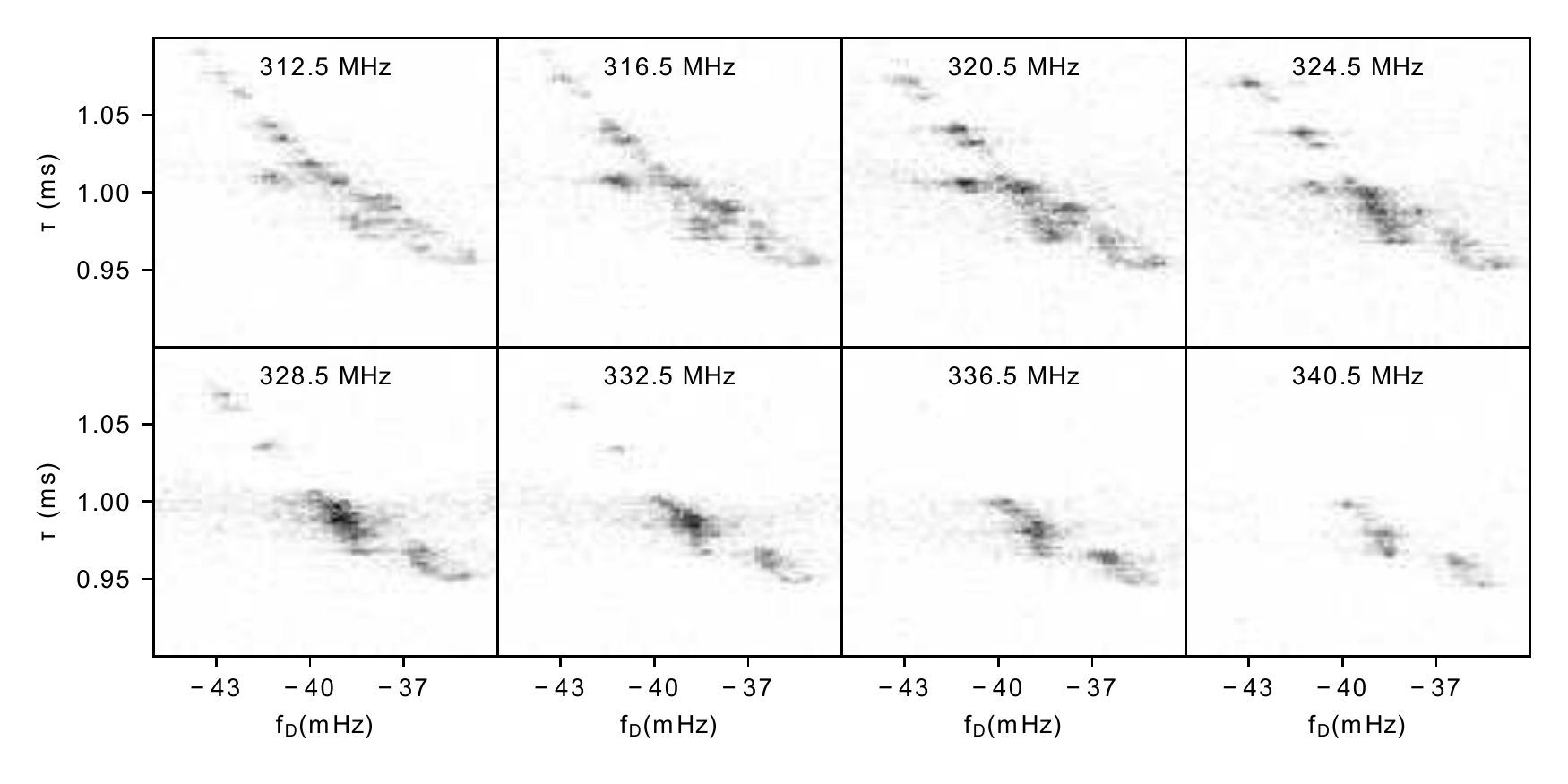}
    \caption{The evolution of the millisecond feature with frequency. Each panel shows the zoomed in of the wavefield's power spectrum for a $4~\rm \text{MHz}$ band with frequency increasing left to right and top to bottom from $312.5~\rm \text{MHz}$ to $340.5~\rm \text{MHz}$. The lower left panel ($328.5~\rm \text{MHz}$) may show evidence of images merging, resulting in the enhanced brightness of the largest feature. The Doppler frequencies have been scaled to keep the curvature of the main arc constant and reduce drift of the millisecond feature.}
    \label{fig:GB057_msf_evo}
\end{figure*}

Using this curvature result, we can now perform the mosaic phase retrieval described in Section~\ref{sec:PhaseRet}. Since our rebinning of the dynamic spectrum removed the millisecond feature, we interpolate our recovered field to the original resolution before applying amplitudes from the dynamic spectrum. Effectively, this deconvolves the conjugate spectrum by the main arc and allows us to see features that were not part of the original one-dimensional model. Fig.~\ref{fig:GB057_wavefield} shows this deconvolved spectrum. The narrow main arc, including features above the delay cutoff imposed by our rebinning, suggests a highly one-dimensional structure. \cite{Brisken_2009} put the axial ratio of this main arc at $R_{ap}>27$ from their VLBI analysis.  However, several deviations from this structure, such as the millisecond feature and an additional feature offset from the main arc near $-24~\rm \text{mHz}$ or $300~\rm \mu \text{s}$, are also visible after using the amplitude measurement from the dynamic spectrum. The millisecond feature is particularly well recovered, with multiple distinct images that can be seen to evolve with frequency in Fig.~\ref{fig:GB057_msf_evo}. The brightening of the central images in the lower left panel may indicate that multiple images are merging as frequency increases, though we leave a detailed analysis for a future work.

{}

\section{Ramifications}\label{sec:conclusion}
In this work, we present a technique for precisely measuring the arc curvature in pulsar secondary spectra via the $\theta-\theta$ transformation. By making use of the full phase information of the conjugate spectrum, as well as the shape of inverted arclets, we are able to improve on traditional methods by orders of magnitude. Measurement of arc curvatures, or equivalently the scintillation timescale, provides a probe for the transverse velocities of the pulsar and interstellar screen. These, in turn, have been used previously to measure inclinations, including their sense, in the double pulsar (PSR J0737$-$3039) by \cite{Rickett_2014} and in PSR J0437$-$4715 by \cite{ReardonPhD}. Other orbital parameters, such as the advance of periastron, are also measurable in this way \cite{Reardon_2019}.

Measuring inclinations is expected to be particularly interesting for certain black widow and redback systems which may have exceptionally high masses. \cite{van_Kerkwijk_2011} used light curve modelling of the companion of PSR B1957+20 to infer a pulsar mass of $M_p = 2.40\pm0.12\rm~M_\odot$. However, difficulties in modelling the companion atmosphere lead to systematic uncertainty in the inclinations measured this way, which allow the mass to be as low as $1.66\rm~M_\odot$. Due to the low relative orbital velocity of the system, traditional methods of measuring the scintillation timescale or curvature changes over the orbit have not yielded results. A $\theta-\theta$ style analysis may be able to detect these variations and provide an inclination.

Alternatively, by including information about $\sin(i)$, the variations in curvature over the Earth and pulsar orbits can be used to measure pulsar distances. If one were to improve distance measurements to less than the wavelength of a gravitational wave, PTAs could make use of the 'pulsar term' in the signal, which can improve signal strength by a factor of two and lead to order of magnitude improvements in localizing sources (\cite{corbin2010pulsar} and \cite{Lee-2011}).

In addition, this method provides an optimal solution for one-dimensional phase retrieval or holography of the interstellar lens. This provides the relative phases, amplitudes and delays of the scattered images. Variations in the scattering delay can be observed on the scale of months to years when measured from the secondary spectrum, and may be underestimated when derived from scintillation bandwidth alone \cite{Main_2020}. Directly determining the delay for each image, may then be an important tool for removing systematic errors from PTAs. Furthermore, performing phase retrieval on VLBI data allows for imaging of the scattering medium as seen in ~\cite{Brisken_2009} and ~\cite{Pen_2014}. Previously, phase retrieval has been achieved through iteratively adding images to a model conjugate wavefield (\cite{Walker2005}), coherently stacking inverted arclets (\cite{Pen_2014}), and cyclic spectroscopy (\cite{Walker_2013}). However, these methods have only been successfully applied to a few systems. $\theta-\theta$ methods will be useful in expanding the number of systems that can be analysed. Though the basic approach of $\theta-\theta$ is based around the assumption of a one dimensional collection of images, it can still provide insight in more complicated cases. For systems similar to B0834+06, where the dominant linear screen is accompanied by a second offset feature, deconvolving the secondary spectrum using our one dimensional model can still isolate images from other structures. For two dimensional collections of images with high aspect ratios, using smaller chunk sizes can reduce our resolution in the secondary spectrum and may produce an effectively one dimensional problem. However, for more complicated lensing examples additional techniques will be required.

\section*{Acknowledgements}
We acknowledge the support of the Natural Sciences and Engineering Research Council of Canada (NSERC), [funding reference number RGPIN-2019-067, 523638-201]. We receive support from Ontario Research Fund—research Excellence Program (ORF-RE), Simons Foundation, Canadian Institute for Advanced Research (CIFAR), and Alexander von Humboldt Foundation. The Arecibo Observatory  is  operated  by  SRI  International  under  a  cooperative agreement with the National Science Foundation (AST-1100968), and in alliance with Ana G. M\'endez-Universidad Metropolitana,  and  the  Universities  Space  Research  Association.  The  National  Radio  Astronomy  Observatory  is  a facility of the National Science Foundation operated under cooperative agreement by Associated Universities, Inc. Computations were performed on the Niagara supercomputer at the SciNet HPC Consortium. SciNet is funded by: the Canada Foundation for Innovation; the Government of Ontario; Ontario Research Fund - Research Excellence; and the University of Toronto. This research made use of Astropy,\footnote{http://www.astropy.org} a community-developed core Python package for Astronomy \cite{astropy:2013, astropy:2018}. Tim Sprenger is a member of the International Max Planck Research School for Astronomy and Astrophysics at the Universities of Bonn and Cologne.

\section*{Data Availability}


The code used in generating $\theta-\theta$ diagrams as well as fitting for curvatures can be found as part of the scintools python package developed by Daniel Reardon at \href{https://github.com/danielreardon/scintools}{github.com/danielreardon/scintools} (\cite{reardon2020precision}).


\bibliographystyle{mnras}
\bibliography{thth} 

\begin{thebibliography}{}
\makeatletter
\relax
\def\mn@urlcharsother{\let\do\@makeother \do\$\do\&\do\#\do\^\do\_\do\%\do\~}
\def\mn@doi{\begingroup\mn@urlcharsother \@ifnextchar [ {\mn@doi@}
  {\mn@doi@[]}}
\def\mn@doi@[#1]#2{\def\@tempa{#1}\ifx\@tempa\@empty \href
  {http://dx.doi.org/#2} {doi:#2}\else \href {http://dx.doi.org/#2} {#1}\fi
  \endgroup}
\def\mn@eprint#1#2{\mn@eprint@#1:#2::\@nil}
\def\mn@eprint@arXiv#1{\href {http://arxiv.org/abs/#1} {{\tt arXiv:#1}}}
\def\mn@eprint@dblp#1{\href {http://dblp.uni-trier.de/rec/bibtex/#1.xml}
  {dblp:#1}}
\def\mn@eprint@#1:#2:#3:#4\@nil{\def\@tempa {#1}\def\@tempb {#2}\def\@tempc
  {#3}\ifx \@tempc \@empty \let \@tempc \@tempb \let \@tempb \@tempa \fi \ifx
  \@tempb \@empty \def\@tempb {arXiv}\fi \@ifundefined
  {mn@eprint@\@tempb}{\@tempb:\@tempc}{\expandafter \expandafter \csname
  mn@eprint@\@tempb\endcsname \expandafter{\@tempc}}}

\bibitem[\protect\citeauthoryear{Aggarwal et~al.}{Aggarwal
  et~al.}{2019}]{Aggarwal_2018}
Aggarwal K.,  et~al., 2019, \mn@doi [Astrophys. J.] {10.3847/1538-4357/ab2236},
  880, 2

\bibitem[\protect\citeauthoryear{{Astropy Collaboration} et~al.,}{{Astropy
  Collaboration} et~al.}{2013}]{astropy:2013}
{Astropy Collaboration} et~al., 2013, \mn@doi [\aap]
  {10.1051/0004-6361/201322068}, \href
  {http://adsabs.harvard.edu/abs/2013A%26A...558A..33A} {558, A33}

\bibitem[\protect\citeauthoryear{{Boyle} \& {Pen}}{{Boyle} \&
  {Pen}}{2012}]{Boyle2012}
{Boyle} L.,  {Pen} U.-L.,  2012, \mn@doi [\prd] {10.1103/PhysRevD.86.124028},
  \href {https://ui.adsabs.harvard.edu/abs/2012PhRvD..86l4028B} {86, 124028}

\bibitem[\protect\citeauthoryear{{Brisken}, {Macquart}, {Gao}, {Rickett},
  {Coles}, {Deller}, {Tingay}  \& {West}}{{Brisken}
  et~al.}{2010}]{Brisken_2009}
{Brisken} W.~F.,  {Macquart} J.~P.,  {Gao} J.~J.,  {Rickett} B.~J.,  {Coles}
  W.~A.,  {Deller} A.~T.,  {Tingay} S.~J.,   {West} C.~J.,  2010, \mn@doi
  [\apj] {10.1088/0004-637X/708/1/232}, \href
  {https://ui.adsabs.harvard.edu/abs/2010ApJ...708..232B} {708, 232}

\bibitem[\protect\citeauthoryear{Corbin \& Cornish}{Corbin \&
  Cornish}{2010}]{corbin2010pulsar}
Corbin V.,  Cornish N.~J.,  2010, Pulsar Timing Array Observations of Massive
  Black Hole Binaries (\mn@eprint {arXiv} {1008.1782})

\bibitem[\protect\citeauthoryear{Lee, Wex, Kramer, Stappers, Bassa, Janssen,
  Karuppusamy  \& Smits}{Lee et~al.}{2011}]{Lee-2011}
Lee K.~J.,  Wex N.,  Kramer M.,  Stappers B.~W.,  Bassa C.~G.,  Janssen G.~H.,
  Karuppusamy R.,   Smits R.,  2011, \mn@doi [Monthly Notices of the Royal
  Astronomical Society] {10.1111/j.1365-2966.2011.18622.x}, 414, 3251

\bibitem[\protect\citeauthoryear{{Lyne} \& {Smith}}{{Lyne} \&
  {Smith}}{1982}]{Lyne1982}
{Lyne} A.~G.,  {Smith} F.~G.,  1982, \mn@doi [\nat] {10.1038/298825a0}, \href
  {https://ui.adsabs.harvard.edu/abs/1982Natur.298..825L} {298, 825}

\bibitem[\protect\citeauthoryear{Main et~al.,}{Main et~al.}{2020}]{Main_2020}
Main R.~A.,  et~al., 2020, \mn@doi [Monthly Notices of the Royal Astronomical
  Society] {10.1093/mnras/staa2955}, 499, 1468

\bibitem[\protect\citeauthoryear{Pen, Macquart, Deller  \& Brisken}{Pen
  et~al.}{2014}]{Pen_2014}
Pen U.-L.,  Macquart J.-P.,  Deller A.~T.,   Brisken W.,  2014, \mn@doi [MNRAS]
  {10.1093/mnrasl/slu010}, 440, L36–L40

\bibitem[\protect\citeauthoryear{{Price-Whelan} et~al.,}{{Price-Whelan}
  et~al.}{2018}]{astropy:2018}
{Price-Whelan} A.~M.,  et~al., 2018, \mn@doi [\aj] {10.3847/1538-3881/aabc4f},
  \href {https://ui.adsabs.harvard.edu/#abs/2018AJ....156..123T} {156, 123}

\bibitem[\protect\citeauthoryear{Putney \& Stinebring}{Putney \&
  Stinebring}{2006}]{Putney_2006}
Putney M.~L.,  Stinebring D.~R.,  2006, \mn@doi [Chinese Journal of Astronomy
  and Astrophysics] {10.1088/1009-9271/6/s2/43}, 6, 233

\bibitem[\protect\citeauthoryear{{Reardon}}{{Reardon}}{2018}]{ReardonPhD}
{Reardon} D.~J.,  2018, PhD thesis, Monash Centre for Astrophysics (MoCA),
  School of Physics and Astronomy, Monash University, Victoria 3800, Australia

\bibitem[\protect\citeauthoryear{Reardon, Coles, Hobbs, Ord, Kerr, Bailes, Bhat
   \& Venkatraman~Krishnan}{Reardon et~al.}{2019}]{Reardon_2019}
Reardon D.~J.,  Coles W.~A.,  Hobbs G.,  Ord S.,  Kerr M.,  Bailes M.,  Bhat N.
  D.~R.,   Venkatraman~Krishnan V.,  2019, \mn@doi [MNRAS]
  {10.1093/mnras/stz643}, 485, 4389–4403

\bibitem[\protect\citeauthoryear{Reardon et~al.,}{Reardon
  et~al.}{2020}]{reardon2020precision}
Reardon D.~J.,  et~al., 2020, \mn@doi [ApJ] {10.3847/1538-4357/abbd40}, 904,
  104

\bibitem[\protect\citeauthoryear{Rickett et~al.,}{Rickett
  et~al.}{2014}]{Rickett_2014}
Rickett B.~J.,  et~al., 2014, \mn@doi [ApJ] {10.1088/0004-637x/787/2/161}, 787,
  161

\bibitem[\protect\citeauthoryear{{Sprenger}, {Wucknitz}, {Main}, {Baker}  \&
  {Brisken}}{{Sprenger} et~al.}{2021}]{Sprenger_2020}
{Sprenger} T.,  {Wucknitz} O.,  {Main} R.,  {Baker} D.,   {Brisken} W.,  2021,
  \mn@doi [\mnras] {10.1093/mnras/staa3353}, \href
  {https://ui.adsabs.harvard.edu/abs/2021MNRAS.500.1114S} {500, 1114}

\bibitem[\protect\citeauthoryear{{Stinebring}, {McLaughlin}, {Cordes},
  {Becker}, {Goodman}, {Kramer}, {Sheckard}  \& {Smith}}{{Stinebring}
  et~al.}{2001}]{Stinebring2001}
{Stinebring} D.~R.,  {McLaughlin} M.~A.,  {Cordes} J.~M.,  {Becker} K.~M.,
  {Goodman} J.~E.~E.,  {Kramer} M.~A.,  {Sheckard} J.~L.,   {Smith} C.~T.,
  2001, \mn@doi [\apjl] {10.1086/319133}, \href
  {https://ui.adsabs.harvard.edu/abs/2001ApJ...549L..97S} {549, L97}

\bibitem[\protect\citeauthoryear{Walker \& Stinebring}{Walker \&
  Stinebring}{2005}]{Walker2005}
Walker M.~A.,  Stinebring D.~R.,  2005, \mn@doi [MNRAS]
  {10.1111/j.1365-2966.2005.09396.x}, 362, 1279

\bibitem[\protect\citeauthoryear{Walker, Melrose, Stinebring  \& Zhang}{Walker
  et~al.}{2004}]{Walker_2004}
Walker M.~A.,  Melrose D.~B.,  Stinebring D.~R.,   Zhang C.~M.,  2004, \mn@doi
  [Monthly Notices of the Royal Astronomical Society]
  {10.1111/j.1365-2966.2004.08159.x}, 354, 43–54

\bibitem[\protect\citeauthoryear{Walker, Demorest  \& van Straten}{Walker
  et~al.}{2013}]{Walker_2013}
Walker M.~A.,  Demorest P.~B.,   van Straten W.,  2013, \mn@doi [ApJ]
  {10.1088/0004-637x/779/2/99}, 779, 99

\bibitem[\protect\citeauthoryear{van Kerkwijk, Breton  \& Kulkarni}{van
  Kerkwijk et~al.}{2011}]{van_Kerkwijk_2011}
van Kerkwijk M.~H.,  Breton R.~P.,   Kulkarni S.~R.,  2011, \mn@doi [ApJ]
  {10.1088/0004-637x/728/2/95}, 728, 95

\makeatother
\end{thebibliography}




\appendix


\bsp	
\label{lastpage}
\end{document}